\input jnl
\input reforder
\input psfig
\def\sech{\hbox{sech}}
\def\frac#1#2{{#1 \over #2}}

\title Optical conductivity
 associated with solitons in the Peierls state 
as modified by zero-point-motion disorder

\author Kihong Kim$^*$
\affil  Physics Department, Ajou University, Suwon 442-749, Korea
\smallskip 
\bigskip 
\centerline{and}
\author John W.~Wilkins
\affil  Physics Department, Ohio State University, Columbus, OH 43210-1168

\abstract We extend previous work to consider the effect of the soliton
on the density of states and conductivity of quasi-one-dimensional
Peierls systems with quantum lattice fluctuations, modeled by a random
static disorder.  Two features have been verified over an order of
magnitude variation in the disorder. (1) The soliton density of states
and the leading edges of both the soliton-to-band and the band-to-band
conductivities have universal scaling forms.  (2) The soliton-to-band
conductivity has the remarkable feature that the leading edge is
accurately predicted by the joint density of states while the trailing
edge tracks the rigid-lattice conductivity.  Or, in other words,
disorder dominates the leading edge, while matrix element effects are
predominant for the trailing edge.

\endtopmatter

\head{I. Introduction}

This paper is the next logical step in working out the consequences of
a simple observation for most quasi-one-dimensional systems:  {\sl the
lattice zero-point motion is comparable to the lattice dimerization}
associated with the Peierls ground state.\refto{mw} The treatment of
quantum lattice fluctuations about the dimerized ground state has to
date resisted theoretical solution.  Below we describe what has been
accomplished.

The {\sl first step} was to model the quantum fluctuations with a static
random potential.\refto{mw}  This proved possible, for temperatures
below the transition temperature, since (i) the characteristic
phonon energies are less than the Peierls energy gap parameter $\Delta$
and (ii) the phonons are then effectively dispersionless.  At the
mathematical level, the gap parameter is supplemented by a complex
random variable $\delta \Delta(x)$ with a zero average and
$$\frac{\langle\delta\Delta(x)\delta\Delta^*(y)\rangle}{\Delta^2} \equiv 
\eta~\delta((x-y)/\xi_o).\eqno(1.1)$$ 
By measuring length in terms of the coherence length 
$\xi_o \equiv\hbar v_F/\Delta$, we can characterize the static
disorder by the dimensionless number $\eta.$ 

This model can be fitted into the standard continuum form for the
electronic part of the Peierls-Fr\" olich Hamiltonian\refto{bz,tlm} (see
Section III). The density of states $\rho_{disorder}$ for the resulting
problem can be solved analytically, either for the case of an
incommensurate lattice distortion\refto{gdad} or for the commensurate
case of a half-filled band.\refto{oefh}  Both cases give similar
results for weak disorder but differ qualitatively for the
experimentally uninteresting case of strong disorder ($\eta > 1$).

In comparison the density of states for the rigid-lattice
$$\rho_{rigid-lattice}(E)=\frac{1}{\pi\xi_o}\frac{1}{\sqrt{E^2-\Delta^2}}
\frac{|E|}{\Delta}\theta(|E|-\Delta),
\eqno(1.2)$$ 
with a gap for energies $|E| < \Delta$ 
has two important features: (1) the square-root
singularity at $E=\pm\Delta$ is smeared out and (2) the density of
states is now finite for all energies and has a substantial subgap
tail.

The {\sl second step} was to extend model (described in detail in
Section III) calculations to compute the optical conductivity.   For
the rigid-lattice the gap in the density of states leads naturally to a
gap in the optical conductivity until $\hbar\omega=2\Delta$ where there is
a square-root singularity.  In contrast all experiments show a
strongly rounded peak about $2\Delta$, and it was known\refto{sy} that
this shape could not be fitted by just rounding the rigid-lattice
results.

A complex analytical and numerical calculation\refto{kmw} leads to a
rounded peak conductivity which resembled existing experimental data.
A striking feature was that the conductivity below the absorption peak
could be fitted with a universal form\refto{fit} which could be used to
deduce the Fermi velocity and $\eta$ from the data.
This work has encouraged more data from the experimentalists now able to
fit their data.\refto{dgkmw,dtagmkl,more}

The {\sl third step} -- in this paper -- is to compute the effect of
solitons on the optical conductivity of quasi-one-dimensional compounds.
Since solitons are the most studied and discussed feature of these
materials, this work is long overdue.  As Section III shows a
considerable enhancement of the techniques developed before\refto{kmw}
was required.  The next section summarizes the principal conclusions.
More extensive results are given in Section IV followed
by conclusions.

\head{II. Summary of Results}

\subhead{A. Review of soliton absorption in rigid-lattice system}

We use this section both to define notation and to describe the physics
of the rigid lattice in a form appropriate for comparing with our
results for disordered systems.  As Heeger \etal\refto{hkss} observed a
midgap absorption is a universal feature of actual Peierls systems,
since solitons can be created by the formation of the material (\eg in
isomerization of {\sl cis} to {\sl trans} polyacetylene) or, more
usually, by doping the material.

The soliton is located midgap with a wavefunction proportional to 
$1/\cosh(x/\xi_o)$
centered in a CDW strand ($-L<x<L$) [$\xi_o \equiv\hbar v_F/\Delta$]. 
 The local density of states is
$$\eqalignno{\rho_{rigid-lattice}^{soliton}(E,x)&= 
\frac{\delta(E)}{\xi_o}\frac{1}{2\cosh^2(x/\xi_o)}\cr 
&+ \frac{1}{\pi\xi_o}\frac{1}{\sqrt{E^2-\Delta^2}}
\left(\frac{|E|}{\Delta} - 
\frac{\Delta}{|E|}\frac{1}{2 \cosh^2(x/\xi_o)}\right)
\theta(|E|-\Delta).&(2.1)\cr}$$
The density of states in the absence of the soliton is given by omitting
both terms with $\cosh^2(x/\xi_o)$ in (2.1). (Note: 
$\int_{-\infty}^\infty dx~\sech^2(x)/2=1.$)

Using this form we could compute the local joint density of states
which would indicate an absorption due to the soliton starting at
$\Delta$ and localized in the vicinity of the soliton.  More accurately
one can compute the conductivity from the golden rule.\refto{fbc} In
particular the matrix element is proportional to the overlap of the
continuum state and the soliton wavefunction:
$$\int \frac{dx}{\xi_o}~e^{ikx}\sech(x/\xi_o) = \pi\sech(\pi
k\xi_o/2)\equiv \sqrt{2}\psi_k.\eqno(2.2)$$ 
Using the excitation spectrum above
the gap $E(k)=\sqrt{\Delta^2+(\hbar v_F)^2k^2}$ we write
$$\eqalignno{
\sigma^{soliton}_{rigid-lattice}(\omega)\bigg|_{soliton-to-band}
 \propto&\int{dk}|\psi_k|^2
\delta(\hbar\omega-E_k)&(2.3)\cr
= \frac{\xi_o}{2L}
\frac{e^2}{\hbar}\frac{\xi_o}{A}
\frac{\pi^2}{2}
&\frac{1}{\sqrt{(\hbar\omega)^2/\Delta^2-1} }
\frac{1}{\cosh^2(\pi\sqrt{(\hbar\omega)^2/\Delta^2-1}/2)}.&(2.4)\cr
}$$
In the last line we give a dimensionally correct form by including
the ``cross-sectional area'' $A$ of the one-dimensional strand.  For reference we note
that the integrated weight of the soliton conductivity is 2.83
$(e^2/\hbar)(\xi_o^2/(2LA))$. 
Most importantly, the conductivity is proportional to the 
density of solitons $\xi_o/(2L)$. 

The dashed line in Figure 1 shows the soliton-to-band conductivity. 
We note for reference there is no analytic form for the band-to-band
conductivity.  Following many workers\refto{other} we use an approximate form
for this conductivity
$$\sigma(\omega)_{rigid-lattice}^{soliton}\bigg|_{band-to-band}= 
\frac{e^2}{\hbar}\frac{\xi_o}{A}
\left(\frac{2\Delta}{\hbar\omega}\right)^2
\frac{2\Delta}{\sqrt{(\hbar\omega)^2-(2\Delta)^2}}
\left(1 -  \frac{\xi_o}{L} \right).  \eqno(2.5)$$
Equations (2.4) and (2.5) will be used for comparion with our results.

\subhead{B. Sample result for absorption in disordered system}

As discussed in the introduction, the effect of the zero-point motion
is approximated by a static disorder which is characterized by a
dimensionless parameter $\eta$.  In section III we will give the
details of the procedure by which we compute the disordered case.  The
important point is that the calculation gives results which could not
be anticipated by just smearing the rigid-lattice case -- that is, with
no disorder.  As noted in the introduction, in the calculation of CDW
conductivity,\refto{kmw} such a convolution gives a large low frequency
conductivity that is not observed and does not even reproduce the
qualitative shape of the frequency-dependent conductivity seen
experimentally.

Figure 2a shows the soliton density of states for rigid-lattice and the
disordered-lattice with a range of $\eta$'s.  In Section IV we will discuss curves
for a variety of $\eta$ and show how the soliton peak can be scaled to
deduce the dependence of the width on $\eta$.  Here we concentrate on
the main features.  The most important feature is the delta function
midgap state (2.1) is broadened.  This broadening has two consequences:
(1) the soliton-to-band absorption will certainly be broadened on the
leading edge and (2) the conductivity will be affected by the doping or
Fermi level.  When the Fermi level is tied to the middle of the soliton
peak, only filled soliton states below the Fermi level can have transitions 
to the empty conduction states.  On the other hand, if the Fermi level 
lies wholly above the soliton band, then the whole soliton band can have
transitions to the conduction band, thus leading to a broader absorption
than when the Fermi level is pinned to the midgap.  This effect 
is real and should be observable.  It is discussed in detail in Section
IV.B.

In Figure 1 the solid line shows the conductivity for the disordered
lattice with the same value of $\eta=0.01$ as for the narrowest 
density of states
in Fig.~2.  Comparing with the dashed curve for the rigid-lattice
we see that the leading edge is broadened as expected.  In contrast
the trailing edge is not appreciably broadened relative to the
rigid-lattice case.  We will return to this in the next subsection.
In section IV there are more detailed comparisons.

\subhead{C. Comparison of rigid-lattice and disordered case}

We consider one more curve in Figure 1, the dotted line for the 
the joint density of states
$$\rho_{JDOS}^{soliton}(\omega) = 
\int dE~
\rho^{soliton}_{disorder}(E)~
\rho^{soliton}_{disorder}(E+\hbar\omega).\eqno(2.6)$$

Figure 3 shows the limits of the integration for transitions involving
the soliton are such to permit two sets of transitions:  (a) from the
filled part of soliton band to empty conduction band and (b) from
filled valence band to the empty part of soliton band.  In the case
where the Fermi level lies in the soliton band both transitions are
possible and lead to a broader joint density of states.  In the case
where the Fermi level is wholly above (below) the soliton band only
case a (b) is permitted and leads to a narrower joint density of
states.

The dotted curve in Figure 1 is the joint density of states for
$\eta=0.01$ with the Fermi level pinned to the midgap.  The striking
feature -- nearly invisible -- is that the leading edges of the joint
density of states  are nearly coincident with that of the full
calculation.  This is a common feature for all $\eta$ and all chemical
potentials (c.f. Section IV).  This suggests that the matrix element is
essentially constant for the leading edge.  On the other hand, for the
trailing edge the dotted line lies clearly above the full calculations.
Indeed as noted in the previous section the trailing edge tracks the
rigid-lattice case which is dominated by the matrix element.   In
summary,

\item{1.}The leading edge is dominated by the joint density of states
for the disordered lattice.
\item{2.}The trailing edge is dominated by full conductivity for the
rigid lattice case.

\noindent We have no simple explanation for relatively sudden switch on
at $\hbar\omega = \Delta$ of the matrix element effects and relatively
sudden switch off of the disorder effects.  But we can affirm that it
has been seen in extensive numerical calculations over the parameter
ranges $0.01 < \eta < 0.5$ and $0 < E_F < 0.5~\Delta.$

\head{III. Method}

We consider the standard continuum model of noninteracting electrons in
one dimension with a Peierls ground state.  Our system is described by
a Dirac-type equation for the wave functions
$\psi_1(x)$ and $\psi_2(x)$ for electrons moving with the Fermi
velocity $v_F$ to the right and to the left, respectively, in the interval 
$-L\leq x\leq L$ with a complex random potential $\delta\Delta(x)$:
$$\eqalignno{
\left( \matrix{ 
-i\hbar v_F\displaystyle{\frac{\partial}{\partial x}}
&\Delta(x)+\delta\Delta(x)\cr
\Delta(x)+\delta\Delta^*(x)
& i\hbar v_F\displaystyle{\frac{\partial}{\partial x}}\cr}\right)
& 
\left(\matrix{\psi_1(x)\cr \psi_2(x)\cr}\right)
=E \left(\matrix{\psi_1(x)\cr \psi_2(x)\cr}\right)&(3.1)\cr
\langle\delta\Delta(x)\rangle=\langle\delta\Delta(x)\delta\Delta(y)\rangle=0,
\langle\delta\Delta(x)&\delta\Delta^*(y)\rangle=\eta\Delta^2\delta((x-y)/\xi_o).
&(3.2)\cr}
$$
In terms of the self-consistent disorder-averaged Peierls gap
$\Delta$ in the ground state and the coherence length $\xi_o=\hbar
v_F/\Delta$, the soliton configuration for the spatially-varying gap
function $\Delta(x)$ is
$$
{\frac{\Delta(x)}{\Delta}}=\tanh\left({\frac{x}{\xi_o}}\right)\eqno(3.3)$$ 

We use $G^+$ ($G^-$), the retarded (advanced) $2 \times 2$ matrix 
Green function, to compute all properties of interest.
The density of states is
$$
\rho(E)=\frac{1}{2L}\int_{-L}^L dx~\langle -\frac{1}{\pi}
{\rm Im}\left[ {\rm Tr}~G^+(x,x|E)\right]\rangle.
\eqno(3.4)$$
The real part of the frequency-dependent conductivity for a strand of
cross-sectional area $A$ at low temperatures
($\hbar\omega\gg k_BT$) is
$$\eqalignno{
\sigma(\omega)=&\frac{e^2}{A\hbar}\frac{2}{\pi\hbar\omega}
\int_{-\hbar\omega+E_F}^{E_F}dE~\frac{1}{2L}\int_{-L}^Ldy\cr
&~~~~\int_y^\infty dx~{\rm Re}~[j^{+-}(E+\hbar\omega,E|x,y)
-j^{++}(E+\hbar\omega,E|x,y)].&(3.5)\cr}
$$
The correlation function in (3.5) involves a trace of Green functions
and the Pauli matrix $\sigma_3$
$$
j^{\pm\pm}(E^\prime,E|x,y)=(\hbar v_F)^2\langle{\rm Tr}~[
\sigma_3G^\pm(y,x|E^\prime)\sigma_3G^\pm(x,y|E)]\rangle.
\eqno(3.6)
$$
Finally we compute the conductivity using the approximation $\langle
GG\rangle\approx\langle G\rangle\langle G \rangle$.

Two linearly independent wave functions $\psi$ and $\tilde\psi$ satisfy
the boundary conditions that $\psi(-L)$ and $\tilde\psi(L)$ do not
diverge as $L$ becomes large, whereas $\psi(L)$ and $\tilde\psi(-L)$
diverge.  These $\psi$ and $\tilde\psi$ will be statistically
independent in the limit $L\rightarrow \infty$.\refto{oe} For $x>y$,
$$
G(x,y|E)=\frac{i}{\hbar v_F(\psi_1\tilde\psi_2-\psi_2\tilde\psi_1)}
\left(
\matrix{\tilde\psi_1(x)\psi_2(y)&\tilde\psi_1(x)\psi_1(y)\cr
\tilde\psi_2(x)\psi_2(y)&\tilde\psi_2(x)\psi_1(y) \cr} \right),
\eqno(3.7)$$
where ($\psi_1\tilde\psi_2-\psi_2\tilde\psi_1$) is a constant
independent of $x$.  We obtain $G^+~(G^-)$ by solving for wave
functions with Im$E$ a small positive (negative) number. Then the
denominator can be Taylor-expanded and
$$
G^+(x,y|E)=-\frac{i}{\hbar v_F}\sum_{n=0}^\infty\left(
\matrix{
\hat C_n(x,y)\hat z_n(x)& \hat D_n(x,y)\hat z_n(x)\cr
\hat C_n(x,y)\hat z_{n+1}(x)& \hat D_n(x,y)\hat z_{n+1}(x)
\cr}\right),\eqno(3.8)
$$
where
$$\eqalignno{
\hat C_n(x,y)&= \hat y_n(x)\frac{\psi_2(y)}{\psi_2(x)},~
\hat D_n(x,y)= \hat y_n(x)\frac{\psi_1(y)}{\psi_2(x)},\cr
\hat y_n(x)=&\left[\frac{\psi_1(x)}{\psi_2(x)}\right]^n,~
\hat z_n(x)=\left[\frac{\tilde\psi_2(x)}{\tilde\psi_1(x)}\right]^n.
&(3.9)\cr}
$$

Next, we take the $L\rightarrow\infty$ limit and utilize the
statistical independence of $\psi$ and $\tilde\psi$ to factor the
average of the Green function into products of averages of $\hat C$
($\hat D$) and $\hat z$.  In general, the quantities
$C_n(x,y)\equiv\langle\hat C_n(x,y)\rangle$ and 
$D_n(x,y)\equiv \langle\hat D_n(x,y)\rangle$ are functions of $x$ and $y$, 
and $y_n(x)\equiv \langle\hat y_n(x)\rangle$ and 
$z_n(x)\equiv\langle\hat z_n(x)\rangle$ are functions of $x$.  
Thus the averaged Eq.~(3.8)
can be written in terms of $z_n(x)$, $C_n(x,y)$ and $D_n(x,y)$.  The
following equations for $y_n$, $z_n$ and $C_n$ can be derived from
Eq.~(3.1) using standard Fokker-Planck methods,\refto{lgp} subject to the
conditions $y_0=z_0=1$, $dy_n/dx|_{x=\pm\infty}
=dz_n/dx|_{x=\pm\infty}=0$, and $C_n(x,x)=y_n(x)$, which follow from
Eq.~(3.9) and the fact that the translational invariance has to be
restored far from $x=0$ where the soliton is located:

$$\eqalignno{
\xi_o\frac{dy_n}{dx}
=&2in\frac{E}{\Delta}y_n-in\frac{\Delta(x)}{\Delta}y_{n+1}
-in\frac{\Delta(x)}{\Delta}y_{n-1}-\eta n^2y_n,
&(3.10)\cr
-\xi_o\frac{dz_n}{dx}
=&2in\frac{E}{\Delta}z_n -in\frac{\Delta(x)}{\Delta}z_{n+1}-
in\frac{\Delta(x)}{\Delta}z_{n-1} -\eta n^2z_n,&(3.11)\cr
\xi_o\frac{dC_n(x,y)}{dx}
=&i(2n+1)\frac{E}{\Delta}C_n -i(n+1)\frac{\Delta(x)}{\Delta}C_{n+1}
-in\frac{\Delta(x)}{\Delta}C_{n-1}\cr
&-\eta\left(n^2+n+\frac{1}{2}\right)C_n.&(3.12)\cr}
$$
The function $D_n$ satisfies the same equation as $C_n$ with
the initial condition $D_n(x,x)=y_{n+1}(x)$.

We can relate $y_n$ and $z_n$ in a simple manner. In the soliton case
with $\Delta(x)=-\Delta(-x)$, we can easily prove that
$z_n(x)=(-1)^ny_n(-x)$.  In order to integrate Eq.~(3.10) numerically,
we choose a starting point ($x=-L$) sufficiently far from $x=0$ that
the spatial variation of $y_n$ is negligible. Then $y_n$ satisfy a set
of algebraic equations:
$$
\left(2\frac{E}{\Delta}+i\eta n\right)y_n-y_{n+1}-y_{n-1}=0,
\eqno(3.13)
$$
which we solve by truncating the number of equations.  With the
solutions of Eq.~(3.13) as initial conditions, we integrate Eq.~(3.10)
from $x=-L$ to $+L$.  These results are used as initial conditions for
Eq.~(3.12). After we obtain $y_n(x)$, $z_n(x)$, $C_n(x,y)$ and
$D_n(x,y)$, and therefore $G^+(x,y|E)$ for $x>y$ and $E>0$, we get all
other Green functions using the symmetry relations
$$\eqalignno{
&G_{11}^{\pm}(y,x|E)=G_{22}^{\pm}(x,y|E)=[G_{11}^{\mp}(x,y|E)]^*,\cr
&G_{12}^{\pm}(y,x|E)=G_{12}^{\pm}(x,y|E)=[G_{21}^{\mp}(x,y|E)]^*,\cr
&G_{11}^{\pm}(x,y|-E)=-G_{22}^{\mp}(x,y|E),\cr
&G_{12}^{\pm}(x,y|-E)=G_{21}^{\mp}(x,y|E).&(3.14)\cr}
$$

The key to our method is the fact that in disordered systems, 
$y_n$, $z_n$, $C_n$, and $D_n$ all decay very rapidly as a function
of $n$. To solve Eqs.~(3.10-13) numerically, we truncate the number of 
equations by setting $y_n=z_n=0$ for $n>N$. The results presented
here was obtained for $N=40$, which was sufficiently large to ensure 
convergence. Our system size is 20 coherence lengths, and therefore
the soliton density is $1/(20\xi_o)$. In the numerical three-dimensional
integration of Eq.~(3.5), we used the mesh sizes of $dE=0.025\Delta$,
$dx=0.05\xi_o$, and $dy=0.25\xi_o$. Convergence of the data
with respect to these parameters was also carefully checked.

\head{IV. Results}

While the basic result was presented in Section II, here we present
additional details on pertinent features.
Among these are role of (a) disorder on the density of
states, (b) chemical potential on the conductivity and
(c) disorder on the soliton-induced conductivity.

\subhead{A. Disorder effect on density of states: scaling}

Figure 2a shows the density of states for values of the disorder from
$0.01 < \eta < 0.08.$  The dotted line gives the density of states in
the absence of disorder (Eq. (2.1));  of course the delta function
at $E = 0$ cannot be exhibited.  We see that the curves evolve smoothly
with increasing $\eta$.

Figure 2b shows the scaling of the soliton band:  $\rho(E,\eta)/\rho
(E=0,\eta)$ versus $E/\Gamma(\eta)$ where the half-width $\Gamma(\eta)$
has to be deduced by fitting.  Except for the largest $\eta$ (0.15 and
0.2) all curves lie on a universal shape which is nearly Gaussian, as
evidenced by the fact that $\rho(E=0,\eta)\Gamma(\eta)$ varies less
than one-half percent over the full $\eta$ range.  The insert in Figure 2b
shows the variation of $\Gamma(\eta)$ with $\eta$.  The dotted line
shows the efforts to do a similar scaling for the leading edge of the
band gap.  Within the scaling the two widths are very similar.  

\subhead{B. Chemical potential: effect on conductivity}

Figure 4 shows how the chemical potential can broaden and shift the
conductivity and, in the process, shows that the leading edge is
predicted by the joint density of states while the trailing edge
reflects the conductivity without disorder.  There are two sets of
three curves:  for the chemical potential = 0 and = $0.55~\Delta$.
Each set has a solid line for the full conductivity, the dotted line
for the joint density of states and dashed curve for the rigid-lattice
conductivity.  Figure 3 shows that the effect of moving the chemical
potential from the middle of the soliton band to wholly above it would
broaden the soliton-to-band conductivity.   Measuring at half height
the width from $\hbar\omega = \Delta$, we see the width is roughly
doubled as the chemical potential moves from midgap to fully above the
soliton band.  As striking is how well the joint density of states
mimics the calculated conductivity.  This further demonstrates that the
leading edge of the conductivity bears no effect of any energy
dependent matrix element.  Less impressive is the agreement of the
trailing edge with the conductivity for the soliton without disorder
(dashed line).

Only space constraints prevent us from showing similar curves for a
wide range of $\eta$.  Of course the effect is less dramatic as the
leading edge broadens but the agreement between $\rho_{JDOS}$ with the
leading edge of the conductivity is common feature over the range
$0.01 < \eta < 0.1$.

\subhead{C. Disorder effect on conductivity}

Figure 5 illustrates the conductivity for four $\eta$ values (0.01,
0.02, 0.05 and 0.1).  While the leading edges of the soliton-to-band
and the band-to-band broaden as expected, the trailing edges of both
are essentially unchanged, tracking the conductivity of the
rigid-lattice result, shown by the dotted line.  While the curve is for
the chemical potential $E_F = 0.55~\Delta$, the results for the midgap
chemical potential are similar.  Figure 5b demonstrates that the
leading edges of the soliton-to-band conductivities have a universal
shape when rescaled.  The insert shows the $\eta$ dependence of the
conductivity half-width $\Gamma_c(\eta)$.  A very similar scaling exists for
the width of the leading edge of the band-to-band conductivity (dotted
line in insert).

\head{V. Discussion and Conclusions}

First we should point out all the failings of the calculation.  These
are as follows: (1) We approximate 
$\langle G G\rangle$ = $\langle G\rangle \langle G\rangle$.  While in
principle we know how to overcome this, the set of differential 
equations to be solved is sufficiently complex that we have not undertaken
it.  (2) We have used the rigid-lattice results for the variation in
$\Delta(x)$ for the soliton.  Once we have introduced disorder we should
self-consistently solve the soliton energy and wavefunction.  This
admirable task is also numerically possible, but we have not attempted
it.  (3) We approximate the quantum fluctuation of
the lattice motion by a random static disorder.  We believe this is a
good approximation in the low-temperature regime but have no hard
estimate of the error.  (4) We have ignored any electron-electron
contribution to any of the density of states or conductivities.  We
would dearly love to consider a tractable model that includes them.

There is another problem with our startling result that there is no
energy-dependent  matrix-element effect on the leading edge of the
soliton-to-band conductivity.  It is solely the results of numerical
calculation.  We cannot yet produce an analytic or physical argument.
Given that we have been working of the ramifications of the large-zero
point motion since late 1991, we do not see a way to this clearly
desirable deeper understanding.  We welcome the advice and comments 
of colleagues in our continuing efforts.

Nonetheless we think there is a very interesting result -- the turn off
of the disorder effects and and the turn on of the matrix-element effects
at $\hbar\omega\sim\Delta$ -- which is aptly summarized at the end of
Section II.   We would strongly encourage experimentalists to produce
better data on soliton-to-band conductivity on better samples.  We have
been impressed by the high-quality data\refto{dgkmw, dtagmkl,more} our
earlier work\refto{kmw} on the optical conductivity of the Peierls
state induced.

Finally we suggest one way that offers an approximate approach to a
parameter-free analysis.  In our earlier calculation\refto{kmw} for the
conductivity of the Peierls state, it was possible to extract $\eta$
from the scaled width of the leading edge of the conductivity.  In this
paper we have extended the calculations to smaller $\eta$'s (0.01-0.1)
than in our earlier work (0.1-1.0).  The widths for the leading edge of
the band-to-band conductivity with and without a soliton roughly
agree.  Accordingly the measured band-to-band width
($\Gamma(\eta)/\Delta = 0.83 \eta^{0.60}$) could be used to extract an
$\eta$ that in turn could predict the soliton-to-band conductivity.

This work was supported by the KOSEF (Grant No. 951-0209-030-1), 
Ajou University Research Fund, and the DOE -- Basic Energy 
Sciences, Division of Materials Sciences.  Cray time on the Ohio 
Supercomputer Center is deeply appreciated.

\vfill\eject
\references
$^*$Electronic mail address: khkim@madang.ajou.ac.kr

\refis{hkss}A.~J.~Heeger, S.~Kivelson, J.~R.~Schrieffer and W.-P.~Su,
\rmp 60, 781, 1988.\par

\refis{mw}R.~H.~McKenzie and J.~W.~Wilkins,
\prl 69, 1085, 1992; {\sl Synth.~Met.} {\bf 55-57}, 4296 (1993).\par

\refis{kmw}K.~Kim, R.~H.~McKenzie and J.~W.~Wilkins,
\prl 71, 4015, 1993.\par

\refis{fbc}K.~Fesser, A.~R.~Bishop and D.~K.~Cambell,
\prb 27, 4804, 1983.\par

\refis{other}
N.~Suzuki, M.~Ozaki, S.~Etemad, A.~J.~Heeger and A.~G.~MacDiarmid,
\prl 45, 1209, 1980;
K.~Maki and M.~Nakahara, \prb 23, 5005, 1981; 
J.~T.~Gammel and J.~A.~Krumhansl, \prb 24, 1036, 1981; 
S.~Kivelson, T.~K.~Lee, Y.~R.~Lin-Liu, J.~Peschel and L.~Yu, 
\prb 25, 4172, 1982;
J.~C.~Hicks and A.~L.~Wasserman, \prb 29, 808, 1984.\par

\refis{bz}S.~A.~Brazovskii and I.~E.~Dzyaloshinskii,
{\sl Zh.~Eksp.~Theor.~Fiz.}  {\bf 71}, 2338 (1976) [Sov. Phys. JETP {\bf
44}, 1233 (1976)].\par

\refis{tlm}H.~Takayama, Y.~R.~Lin-Liu and K.~Maki, \prb 21, 2388, 1980.
\par

\refis{oefh}A.~A.~Ovchinnikov and N.~S.~\' Erikhman,
{\sl Zh.~Eksp.~Theor.~Fiz.}  {\bf 73}, 650 (1977) [Sov. Phys. JETP {\bf
46}, 340 (1977)]; H.~J.~Fischbeck and R.~Hayn, {\sl Phys.~Status
Solidi (b)} { \bf 158}, 565 (1990).\par

\refis{oe}A.~A.~Ovchinnikov and N.~S.~\' Erikhman,
{\sl Zh.~Eksp.~Theor.~Fiz.}  {\bf 78}, 1448 (1980) [Sov. Phys. JETP {\bf
51}, 728 (1980)].\par

\refis{gdad}
L.~P.~Gor'kov and O.~I.~Dorokhov, 
{\sl Fiz.~Nizk.~Temp.} {\bf 4}, 332 (1978)
[Sov. J. Low Temp. Phys. {\bf 4}, 160 (1978)];	
A.~Abrikosov and E.~A.~Dorotheyev, {\sl J.~Low Temp.~Phys.} {\bf 46}, 53
(1982).\par

\refis{sy}Z.-B.~Su and L.~Yu, {\sl Comm. Theor. Phys. (Beijing)} {\bf
2}, 1341 (1983).\par

\refis{fit}Below the absorption peak 
$\sigma(\omega_{peak})$, the conductivity can be fitted with the function
$\sigma(\omega)/\sigma(\omega_{peak})=\exp[
-0.49|(\omega-\omega_{peak})/\Gamma(\eta)|^2 
-0.20|(\omega-\omega_{peak})/\Gamma(\eta)|^3]$  and 
$\Gamma(\eta)/\omega_{peak}=\eta^{0.62}(0.414+0.077\eta).$

\refis{dgkmw}
L.~Degiorgi, G.~Gr\"uner, K.~Kim, R.~H.~McKenzie and P.~Wachter,
\prb 49, 14754, 1994.\par

\refis{dtagmkl}
L.~Degiorgi, St.~Thieme, B.~Alavi, G.~Gr\"uner,
 R.~H.~McKenzie, K.~Kim and F.~Levy,
\prb 52, 5603, 1995.\par

\refis{more}K.~A.~Coplin, S.~Jasty, S.~M.~Long, S.~K.~Manohar, Y.~Sun,
A.~G.~MacDiarmid and A.~J.~Epstein, \prl 72, 3206, 1994 [pernigraniline];
N.~Kuroda, M.~Nishida and M.~Yamashita,
private communication [PtCl].\par 

\refis{lgp}I.~M.~Lifshits, S.~A.~Gredeskul and L.~A.~Pastur, {\sl 
Introduction to the Theory of Disordered Systems} (Wiley, New York, 1988),
p.~146.\par

\endreferences

\figurecaptions

\noindent Fig.~1.  Frequency-dependent conductivity for system of one soliton
in a Peierls state for a one-dimensional chain of length $20\xi_o$.
Solid line corresponds to disorder parameter $\eta =0.01,$ while dashed
line to rigid-lattice ($\eta=0$).  Dotted curve is the joint density of
states for $\eta = 0.01$, scaled to match conductivity peak at
$E=\Delta.$  Note leading edge of conductivity is indistinguishable from
joint density of states while trailing edge is similar to rigid-lattice
case.  These two features are seen for all disorder parameters and for
all chemical potentials (doping levels).  In this case, chemical potential
lies in the center of soliton band.  Small structure at $\hbar \omega =
0$ is a  numerical artifact.

\noindent Fig.~2.  Density of states for system of one soliton centered
in $20\xi_o$ chain of Peierls state for disorder parameters $\eta = 0$
(dotted line), 0.01, 0.02, 0.04 and 0.08.  The widths of the
soliton and band density of states  increase with $\eta.$  In (b)
the scaled soliton density of states $\rho(E,\eta)/\rho(E=0,\eta)$ is
plotted versus $E/\Gamma(\eta),$ thus demonstrating the universal shape
for $\eta < 0.1.$  Only large disorder ($\eta$ = 0.15, 0.2) cases
break away from universal shape for $E/\Gamma(\eta) > 2.$  The insert
shows the $\eta$ dependence of $\Gamma(\eta)/\Delta$ for both soliton
band (solid line: $0.465~\eta^{0.49}$) and leading band edge (dotted
line).

\noindent Fig.~3.  Schematic plots of the transitions involved in 
the joint density of states for the $E_F=0$ and $0.55~\Delta.$  The
symbols s-b and b-b correspond to soliton-to-band and band-to-band
transitions.  Since the $E_F=0$ soliton-to-band transition covers only
half the soliton band, that joint density of states will be roughly
half as wide as the $E_F=0.55~\Delta$ curve as reflected in Fig.~3.

\noindent Fig.~4.  Fermi-energy dependence of the soliton-to-band
conductivity for disorder parameter $\eta=0.01$.  For each solid line,
corresponding to $E_F=0$ and $0.55~\Delta,$ there is a nearly
coincident joint density of states for the same chemical potential.
This coincidence is seen for the full range of $\eta$ values.
The rigid-lattice conductivity (dashed line) does not track the trailing
edge of the disordered-lattice conductivity with the same fidelity as
the joint density of states does the leading edge.

\noindent Fig.~5.  Conductivity for disorder parameters $\eta = 0$
(dotted line), 0.01, 0.02, 0.05 and 0.1; all for $E_F= 0.55~\Delta.$
In (a) note that rigid-lattice case (dotted) closely tracks the
trailing edges of all the disordered-lattice cases.  In (b) the scaled
soliton-to-band conductivity $\sigma(\omega)/\sigma_{peak}(\eta)$ is
plotted versus $(\omega-\omega_{peak})/\Gamma_c(\eta),$ thus
demonstrating the universal shape for the leading edge.  The insert
shows the $\eta$ dependence of $\Gamma_c(\eta)/\Delta$ for both
soliton-to-band (solid line: $0.685~\eta^{0.513}$) and band-to-band
(dotted line).
\endfigurecaptions
\vfill\eject

\nopagenumbers
\vbox to 9in{\psfig{file=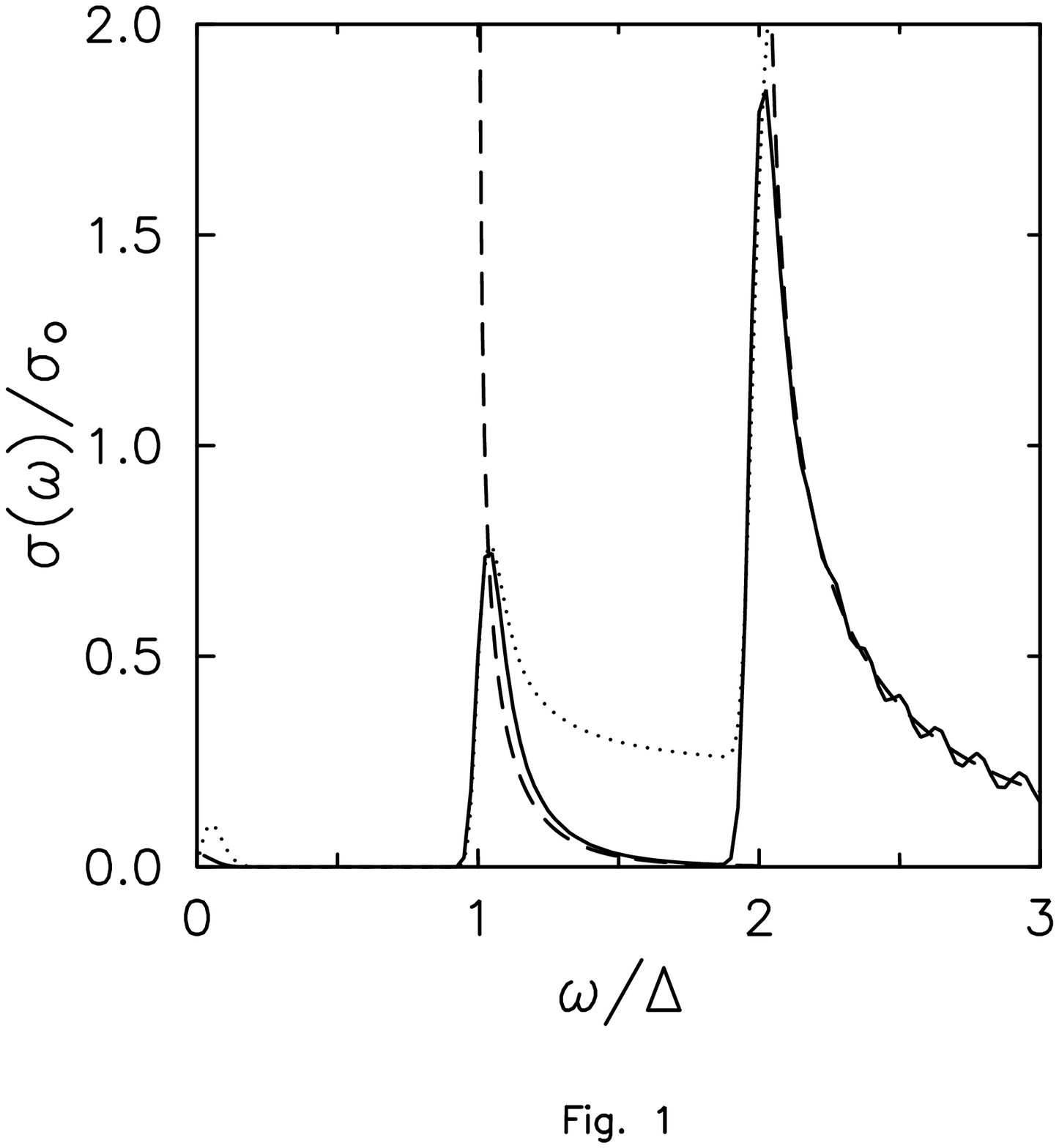}}
\vbox to 9in{\psfig{file=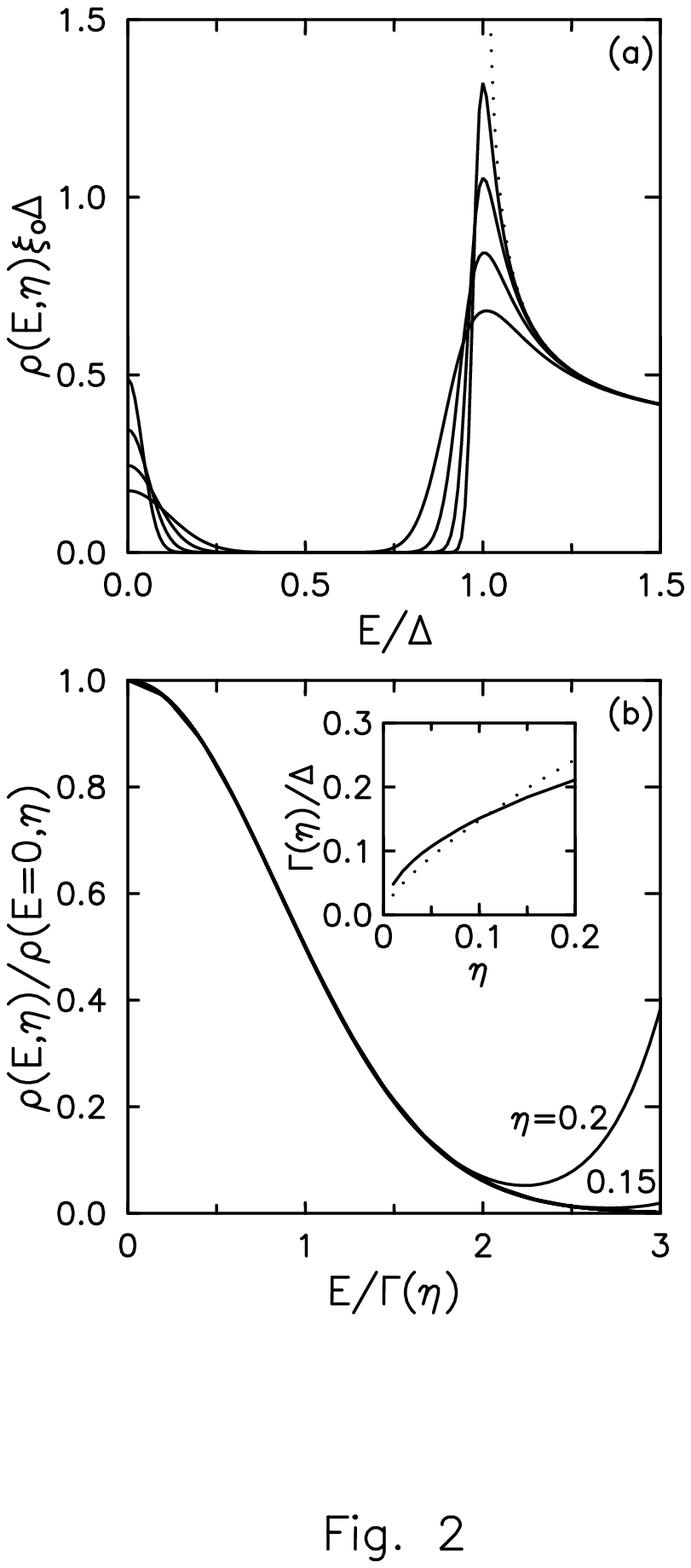}}
\vbox to 9in{\psfig{file=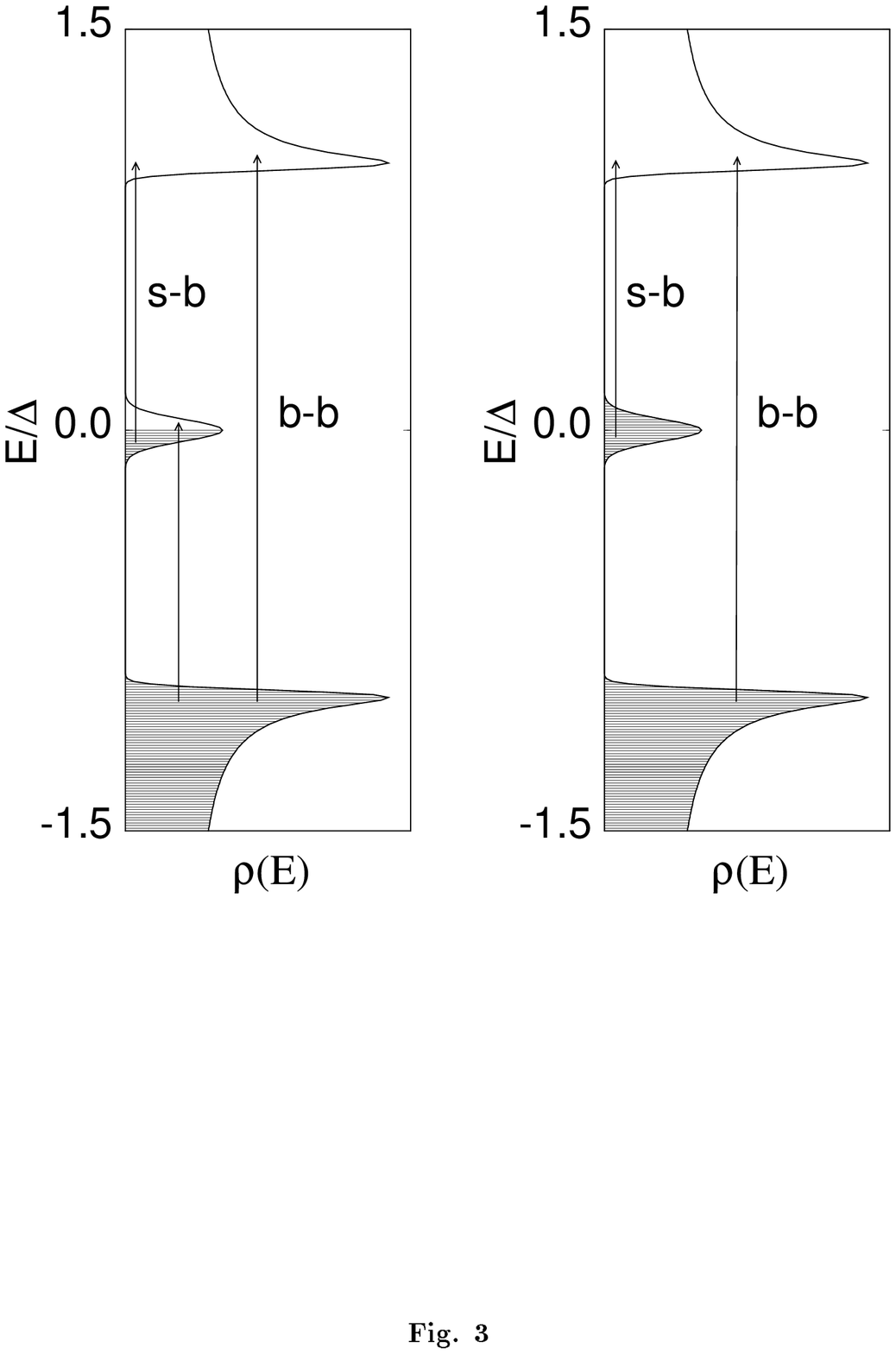}}
\vbox to 9in{\psfig{file=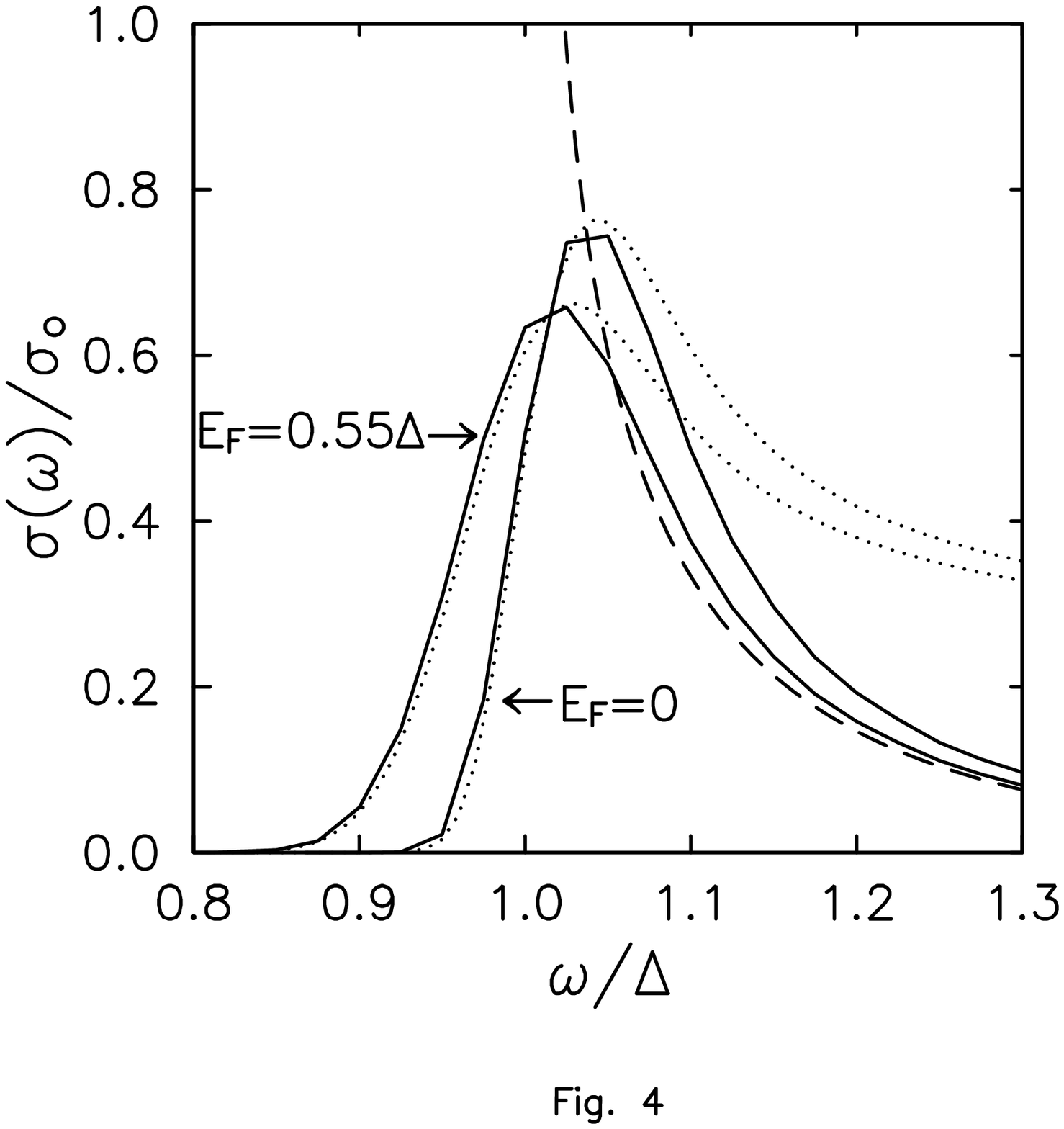}}
\vbox to 9in{\psfig{file=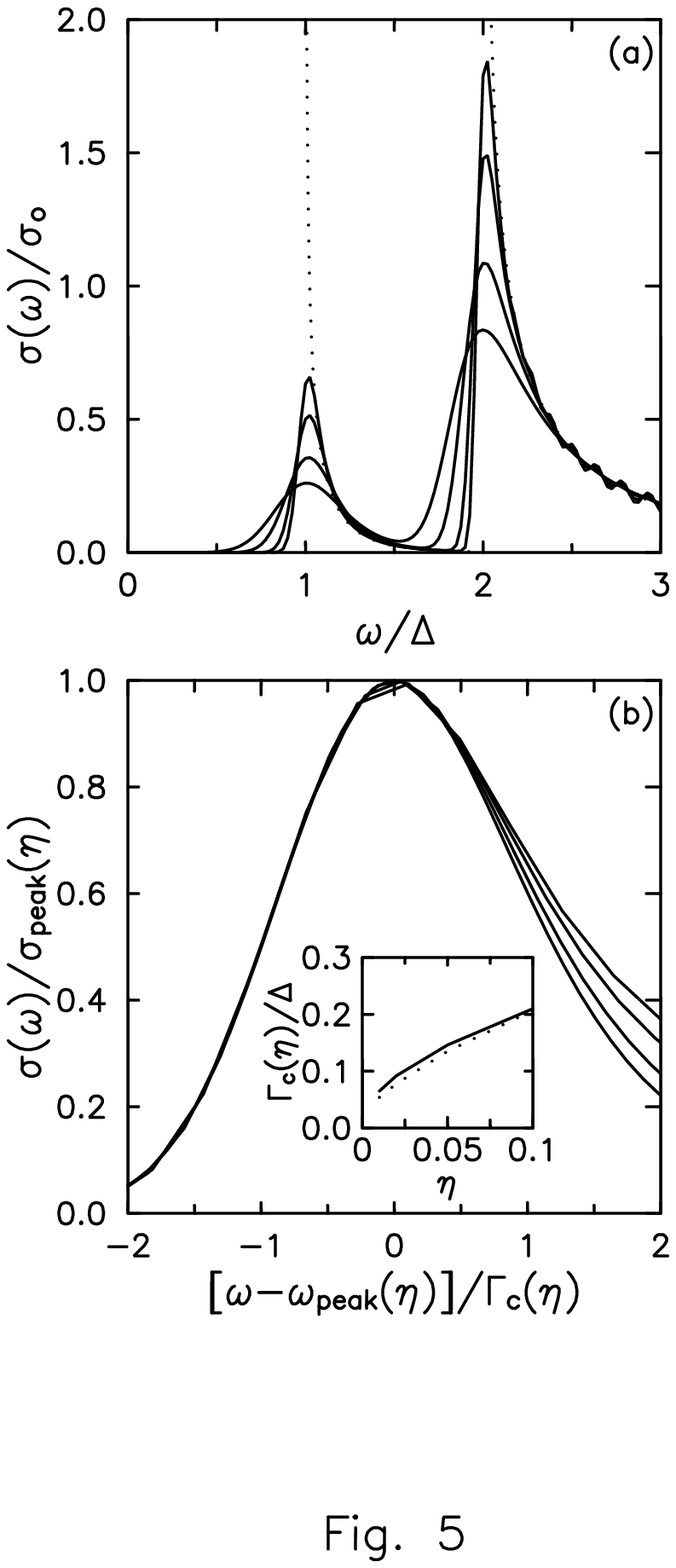}}

\bye